\documentclass[preprint,aps,pra,showpacs,floatfix]{revtex4}
\usepackage{graphicx}
\usepackage{times}
\usepackage{nicefrac}
\usepackage{amsmath}
\usepackage{amsfonts}
\usepackage{amssymb}
\usepackage{amsthm}
\usepackage{epsf}
\usepackage{bm}
\usepackage{bbm}
\usepackage{times}

\usepackage{dcolumn}
\newcolumntype{.}{D{x}{}{-1}}

\newcommand{\hsp}{\hspace}
\newcommand{\be}{\begin{eqnarray}}
\newcommand{\ee}{\end{eqnarray}}

\newcommand{\veps}{\varepsilon}

\newcommand{\vphi}{\varphi}
\newcommand{\balpha}{\bm{\alpha}}
\newcommand{\bnabla}{\bm{\nabla}}

\newcommand{\bmu}{\bm{\mu}}

\newcommand{\bfJ}{{\bf J}}
\newcommand{\bfL}{{\bf L}}
\newcommand{\bfS}{{\bf S}}
\newcommand{\bfT}{{\bf T}}
\newcommand{\rmd}{{\rm d}}
\newcommand{\bfk}{{\bf k}}

\newcommand{\bfr}{{\bf r}}
\newcommand{\bfx}{{\bf x}}
\newcommand{\bfy}{{\bf y}}

\newcommand{\sixj}[6]{
        \left\{
        \begin{array}{ccc}
        #1  & #2  & #3   \\
        #4  & #5  & #6   \\
        \end{array}
        \right\}
        }
\newcommand{\ddd}{\rmd^3}

\newcommand{\az}{\alpha Z}
\newcommand{\mub}{\mu_{\rm B}}
\newcommand{\hD}{h^{\rm D}}
\newcommand{\hDF}{h^{\rm DF}}
\newcommand{\Hnp}{H^{\rm np}}
\newcommand{\Hme}{H}

\newcommand{\tpres}{\tau_{\rm pres}}

\newcommand{\wif}{W^{{i} \to {f}}}
\newcommand{\wifnr}{\wif_{\rm nr}}
\newcommand{\dwif}{\Delta \wif}
\newcommand{\dwifd}{\dwif_{\rm D}}
\newcommand{\dwifci}{\dwif_{\rm CI}}
\newcommand{\dwifneg}{\dwif_{\rm neg}}
\newcommand{\dwifqed}{\dwif_{\rm QED}}
\newcommand{\dwiffreq}{\dwif_{\rm freq}}

\newcommand{\dwd}{\Delta W_{\rm D}}
\newcommand{\dwci}{\Delta W_{\rm CI}}

\newcommand{\dwqed}{\Delta W_{\rm QED}}

\newcommand{\bfTnr}{\bfT_{\rm nr}}

\newcommand{\IC}{I_{\rm C}}

\newcommand{\Psiopt}{\Psi_{\rm opt}}

\newcommand{\jji}{J_{i}}
\newcommand{\mmi}{M_{i}}
\newcommand{\eei}{E_{i}}
\newcommand{\jjf}{J_{f}}
\newcommand{\mmf}{M_{f}}
\newcommand{\eef}{E_{f}}

\newcommand{\braf}{\langle{f}\mid}
\newcommand{\keti}{\mid{i}\rangle}

\newcommand{\brafr}{\langle \jjf \parallel}
\newcommand{\ketir}{\parallel \jji \rangle}

\begin{document}

\title{Magnetic-dipole transition probabilities
in B-like and Be-like ions}
\author{I. I. Tupitsyn,$^1$ A. V. Volotka,$^{1,2}$
D. A. Glazov,$^{1}$ V. M. Shabaev,$^{1,3}$\\
G. Plunien,$^2$ J. R. Crespo L\'opez-Urrutia,$^4$ 
A. Lapierre,$^4$ and J. Ullrich$^4$}

\affiliation{
$^1$ Department of Physics, St. Petersburg State University, Oulianovskaya 1,
Petrodvorets, 198504 St. Petersburg, Russia\\
$^2$ Institut f\"ur Theoretische Physik, TU Dresden,
Mommsenstra{\ss}e 13, D-01062 Dresden, Germany \\
$^3$  Max-Planck Institut f\"ur Physik Komplexer Systeme,
N\"othnitzer Stra{\ss}e 38, D-01187 Dresden, Germany\\
$^4$ Max-Planck Institut f\"ur Kernphysik,
Saupfercheckweg 1, D-69117 Heidelberg, Germany\\
}

\begin{abstract}
The magnetic-dipole transition probabilities between
the fine-structure levels $(1s^2 2s^2 2p) \, ^2P_{1/2} \,-\, ^2P_{3/2}$
for B-like ions and $(1s^2 2s 2p) \, ^3P_{1} \,-\, ^3P_{2}$
for Be-like ions are calculated.
 The configuration-interaction method in the Dirac-Fock-Sturm
basis is employed for the evaluation of the interelectronic-interaction
correction with negative-continuum spectrum being taken into account. 
The $1/Z$
interelectronic-interaction contribution is 
derived within a rigorous QED approach
employing the two-time Green function method. 
The one-electron QED correction is evaluated within
framework of the anomalous magnetic-moment approximation.
A comparison with the theoretical results of other authors and
with available experimental data is presented.
\end{abstract}


\maketitle

%
\section{Introduction}
During the last years, the precision of measurements of
magnetic-dipole (M1) transitions between the fine-structure levels
in highly charged ions has been continuously increased
\cite{Back,Moehs,Trabert1,Trabert2,k15,ti17,Lapierre,Lapierre1}.
Since in some cases the M1 transitions are sufficiently sensitive to
relativistic-correlation and quantum-electrodynamic (QED) effects,
this provides good prospects for probing their influences
on atomic transition probabilities.

To date, a vast number of theoretical calculations of M1-transition probabilities
between the fine-structure levels in highly charged ions has been performed
(see, e.g., Refs. \cite{cheng79,fisch83,cha01}). However, none of these works
have provided a systematic analysis of various effects on the transition probability.
Such an analysis for the $(1s^2 2s^2 2p) \, ^2P_{1/2} \,-\, ^2P_{3/2}$ transition
in B-like ions and for the $(1s^2 2s 2p) \, ^3P_{1}\,-\, ^3P_{2}$ 
transition in Be-like ions is given in the present paper.

To calculate the decay rate one requires knowledge of the transition
energy and the matrix element of the transition operator. Within this work we
employ experimental values of the transition energy, which are measured
accurately enough for the ions under consideration. 

To analyse the influence of various effects, we decompose the
transition probability $\wif$ into several terms, 
\begin{eqnarray}
  \nonumber \wif \,=\, \wifnr \,+\, \dwifd \,+\, \dwifci\,+\, \dwifneg
  \,+\, \dwifqed \,+\, \dwiffreq \,.
\end{eqnarray}
Here $\wifnr$ represents the nonrelativistic M1-transition probability
derived employing the LS-coupling scheme. Within the LS-coupling scheme,
the amplitude of the magnetic-dipole
transition is nonzero only between the fine-structure levels and
depends on the quantum numbers $L$, $S$, and $J$ of the initial and
the final state \cite{Sobelman}. This implies that the contribution
of the interelectronic-interaction vanishes in the nonrelativistic limit.
The explicit expression for 
$\wifnr$ is presented in Section \ref{sec:mdtp}.

The relativistic correction $\dwifd$ is obtained by employing
the one-electron Dirac wave functions for the initial and the final state.
For the relativistic case the interelectronic-interaction contribution
is nonzero, but it is generally suppressed by a factor $(\az)^2/Z$. 
For instance, in case of B-like Ar it amounts to about $0.1\%$.
The interelectronic-interaction correction is, however, rather important
for the $(1s^2 2s 2p) \, ^3P_{1}\,-\, ^3P_{2}$ transition in Be-like
ions, where the terms $^3P_{1}$ and $^1P_{1}$ are strongly mixed. 
In this investigation two approaches are employed for evaluating
the interelectronic-interaction correction. The first one is based
on the configuration-interaction (CI) method in the Dirac-Fock-Sturm basis,
whereas the second one employs perturbation theory with respect to $1/Z$.
Utilizing the CI method the relativistic Hamiltonian is specified within
the no-pair approximation \cite{Sucher,Mittleman,gla04}.
The corresponding contribution to the M1-transition probability
is denoted by $\dwifci$. The evaluation of this term is described
in Section \ref{sec:iicbr}.

The no-pair Hamiltonian does not account for the
negative-energy excitations in the many-electron wave function.
However, this effect, being dependent on the choice of the
one-electron basis, can become significant
\cite{Indelicato,Derevianko}. In Section \ref{sec:iicneg}, the
contribution due to the negative-spectrum $\dwifneg$ is derived.

In Section \ref{sec:iicho}, the interelectronic-interaction correction
of first order in $1/Z$ is evaluated within a rigorous QED approach
employing the two-time Green function method \cite{Shabaev1}. 
Together with verifying the terms $\dwifci$ and $\dwifneg$
to first order in $1/Z$, this provides
the contribution $\dwiffreq$, which incorporates the $1/Z$ 
interelectronic-interaction corrections of higher orders
in $\az $.

Finally, $\dwifqed$  is the QED correction. The evaluation of
this correction to the lowest orders in $\alpha $ and $\alpha Z$ is
described in Section \ref{sec:qed}.

The main goal of the present work is to evaluate the lifetimes
of the states $(2s^2 2p)$ $^2P_{3/2}$ in B-like ions and $(2s 2p)$ $^3P_{2}$
in Be-like ions to utmost accuracy and to investigate the influence
of various effects on the M1-transition probability. The corresponding
analysis is presented in Section \ref{sec:results}.

Atomic units ($\hbar=e=m=1$) are used throughout the paper.

\section{Magnetic-dipole transition probability}
\label{sec:mdtp}
%
The spontaneous $L$-pole transition probability from the initial
state $i$ to the final state $f$ 
reads \cite{Grant1}
\begin{equation}
  \wif_L \,=\, \frac{2\pi}{2\jji+1} \, \sum_{\mmi} \, \sum_{\mmf} \, \sum_M \,
  |A_{LM}|^2 \,,
\end{equation}
where the initial state has the angular momentum $\jji$,
its $z$-projection $\mmi$, and the energy $\eei$, and 
$\jjf$, $\mmf$, $\eef$ denote the corresponding quantum numbers
and the energy of the final state. The transition amplitude
$A_{LM}$ is defined as
\begin{equation}
  A_{LM} \,=\, i^{L+1} \, \sqrt{\frac{\omega}{\pi c}} \, \sqrt{2L+1} \,
  \braf T^L_M \keti \, .
\end{equation}
Here $T^L_M$ denote the components of the multipole transition operator
$\bfT^L$, which is a spherical tensor of rank $L$. In case of a magnetic
transition, $\bfT^L$ is proportional to the tensor product of the
Dirac-matrix vector $\balpha$ and the spherical tensor
$C^L_M= \sqrt{4\pi/(2L+1)} \, Y_{LM}$ \cite{Grant1}
\begin{equation}
  T^L_M \,=\, -i \, j_L(\omega r/c) \, (\balpha \otimes {\bf C}^L )^L_M \,,
\end{equation}
where $j_L$ is the spherical Bessel function and $\omega = \eei - \eef$
is the frequency of the emitted photon.

The magnetic transition probability can be expressed in terms of
the reduced matrix element of $T^L_M$
\begin{equation}
  \wif_L = \frac{2(2L+1)}{2\jji+1}\frac{\omega}{c} \left|  \langle f
  \parallel {\bf T}^L \parallel i \rangle  \right|^2 \,.
\end{equation}
For the magnetic-dipole transition ($L=1$), the tensor product
can be written in terms of the vector product
\begin{equation}
  \bfT^1 = \frac{1}{\sqrt{2}} \, j_1(\omega r/c) \, \frac{[ \balpha \times \bfr ]}{r}
  \,=\, \frac{\sqrt{2}}{r} \, j_1(\omega r/c) \, \bmu \,,
\label{eq:Tdip}
\end{equation}
where $\bmu = - e\,[\bfr \times \balpha]/2$ is the relativistic magnetic moment
operator. Taking into account the first term in the expansion of
$j_1(\omega r/c)$ only and turning into the nonrelativistic limit,
one derives the following relation
between the M1-transition operator $\bfTnr^1$ and the magnetic
moment operator $\bmu_{\rm nr}$
\begin{equation}
  \bfTnr^1 \,=\, \frac{\sqrt{2}}{3} \, \frac{\omega}{c} \bmu_{\rm nr} \,.
\end{equation}
The nonrelativistic magnetic moment operator is given by
\begin{equation}
  \bmu_{\rm nr} = - \mub \, (\bfL + 2 \bfS) \,,
\end{equation}
where $\bfL$ and $\bfS$ are the orbital and spin angular momentum
operators, respectively, and $\mub = |e| \hbar / 2 m c $ denotes Bohr
magneton.

In the $LS$-coupling scheme, which is realized in the nonrelativistic case,
the magnetic-dipole transition probability is nonzero only between
fine-structure levels with $\Delta J = \pm 1$ \cite{Sobelman}.
The reduced matrix element of $\bfTnr^1$ within the $LS$-coupling
is given by
\begin{equation}
  \langle J_f \parallel \bfTnr^1 \parallel J_i \rangle = - \,
  \frac{\sqrt{2}}{3} \, \frac{\omega}{c}\mub \langle J_f \parallel
  (\bfJ + \bfS) \parallel J_i \rangle = - \, \frac{\sqrt{2}}{3}
  \frac{\omega}{c}\mub \langle J_f \parallel \bfS \parallel J_i \rangle \,.
\end{equation}
Utilizing the general formula for the reduced matrix element of the spin
operator \cite{Varshalovich} yields the corresponding expression
for the transition probability
\begin{equation}
  \wifnr = \frac{4\omega^3}{3c^3} \mub^2 \delta_{L_i,L_f} \delta_{S_i,S_f}
    S_i(S_i+1)(2S_i+1)(2J_f+1) \sixj{S_i}{L_i}{J_i}{J_f}{1}{S_i}^2 \,.
\end{equation}
In particular, for the $2s^2 2p_{3/2} \to 2s^2 2p_{1/2}$ transition
one can easily find
\begin{equation}
  \wifnr = \frac{4\omega^3}{9c^3} \mub^2 =
  \frac{1}{3\lambda^3} \, 2.697 3500 \cdot 10^{13} \, {\rm [s^{-1}]}\,,
\end{equation}
where $\lambda$ is the transition wavelength, in \AA.
Thus, in the nonrelativistic limit the magnetic-dipole transition
probability is completely determined by the quantum numbers of the
initial and final states. 
%

\section{Interelectronic interaction in the Breit approximation}
\label{sec:iicbr}
%
To evaluate the interelectronic-interaction contributions, we start with
the relativistic Hamiltonian in the no-pair approximation,
\begin{equation}
  \Hnp = \Lambda_{+} \Hme \Lambda_{+} \,,
\qquad
  \Hme = \sum_i \hD(i) + \sum_{i < j} V(i,j)\,,
\label{eq:Hnp}
\end{equation}
where $\hD(i)$ is the one-particle Dirac Hamiltonian and
the index $i=1,\ldots ,N$ enumerates the electrons.
The Coulomb-Breit interaction operator
$V(i,j)=V_{\rm C}(i,j)+V_{\rm B}(i,j)$ is
specified in coordinate space as
\begin{equation}
  V_{\rm C}(i,j) = \frac{1}{r_{ij}} \,, \qquad
  V_{\rm B}(i,j) = - \frac{\balpha_i \cdot \balpha_j}{r_{ij}}
    - \frac{1}{2}(\balpha_i \cdot \bnabla_i)
                 (\balpha_j \cdot \bnabla_j) r_{ij} \,.
\label{eq:Hint}
\end{equation}
The frequency-dependent part of the full QED interaction operator,
which is beyond the Breit approximation and gives rise to the terms
of higher orders in $\az$, will be considered in Section \ref{sec:iicho}.
$\Lambda_{+}$ is the projector on the positive-energy states,
which can be represented as the product of the one-electron
projectors $\lambda_{+}(i)$ as
\begin{equation}
  \Lambda_{+} = \lambda_{+}(1) \cdot \cdot \cdot \lambda_{+}(N) \,
\label{eq:Lambda}
\end{equation}
together with
\begin{equation}
  \lambda_{+}(i) = \sum_{n} \mid u_n(i) \rangle \langle u_n(i) \mid \,.
\label{eq:lambda}
\end{equation}
Here $u_n$ are the positive-energy eigenstates of an effective
one-particle Hamiltonian $h^u$
\begin{equation}
h^u \, u_n \,=\, \varepsilon_n \, u_n \,,
\end{equation}
which can be taken to be the Dirac Hamiltonian
$\hD$, the Dirac Hamiltonian in an external field or 
the Hartree-Fock-Dirac Hamiltonian in an external field
\cite{Sucher,Mittleman,gla04}.

In order to determine the space of one-electron functions $\left\{
\varphi_n\right\}_{n = 1}^M$, we employed the combined Dirac-Fock (DF)
and the Dirac-Fock-Sturm (DFS) basis set. Here the index $n$ enumerates
different occupied and vacant one-electron states. For the occupied
atomic shells, the orbitals $\varphi_{n}$ with $n=1,\dots, M_0$ were
obtained by the standard restricted Dirac-Fock (RDF) method,
based on a numerical solution of the radial RDF equations
\cite{Grant2,Bratsev}. Only the Coulomb part $V_{\rm C}(i,j)$ of the
Coulomb-Breit interaction operator (\ref{eq:Hint}) was included in the
RDF Hamiltonian $\hDF$.

The vacant orbitals ${\varphi}_{n}$ with $n=M_0+1,\dots, M$ were
obtained by solving the Dirac-Fock-Sturm equations
\begin{equation}
  \left [ \hDF - \veps_{n_0} \right ] \vphi_{n}
  = \xi_{n} W(r) \vphi_{n}\,,
\label{eq:sturm}
\end{equation}
which can be considered as a generalization of the method proposed in
Ref. \cite{Gruzdev} to the relativistic Hamiltonian and to an arbitrary
constant-sign weight function $W(r)$. For every relativistic quantum
number $\kappa$ we choose an occupied DF function $\varphi_{n_0}$,
which we call as reference DF orbital and $\veps_{n_0}$ in
(\ref{eq:sturm}) is the energy of this orbital. The parameter $\xi_{n}$
in Eq.~(\ref{eq:sturm}) can be considered as an eigenvalue of the
Sturmian operator. Obviously, for $\xi_n=0$ the Sturmian function
coincides with the reference DF orbital $\varphi_{n_0}$. If  $W(r) \to 0$
at $r \to \infty$, all Sturmian functions $\varphi_{n}$ have the same
exponential asymptotics at $r \to \infty$. Therefore, the all set of
eigenfunctions of the Dirac-Fock-Sturm operator forms a discrete set
in the space of one-electron wave functions. The completeness of
this basis in the nonrelativistic limit is well-known fact. In
the relativistic case this problem is more complicated and we examined
the completeness of the pure DFS basis, which we used in our many-electron
atomic calculations, numerically, reproducing exact hydrogenlike wave
functions for the same nuclear charge number $Z$. 
It should be noted that the DFS orbitals are orthogonal with respect
to the weight function $W(r)$ and, therefore, form a linear
independent basis set. The completeness and linear independence of the
combined DF and DFS basis was also examined numerically.

In the nonrelativistic theory the widely used choice of the weight
function is $W(r)=1/r$, which leads to the well-known
``charge quantization''. In the relativistic case, however, this
choice is not very suitable, since the behaviour of the Sturmian wave
functions at the origin differs from that of the Dirac-Fock orbitals.
In our calculations we employed the following weight function
\begin{equation}
  W(r) = \frac{1 - \exp[-(\alpha r)^2]}{(\alpha r)^2} \,,
\label{eq:weight}
\end{equation}
which, unlike $1/r$, is regular at the origin.

To generate the one-electron wave functions $u_n$, we used the
unrestricted DF (UDF) method in the joined DF and DFS basis,
\begin{equation}
  u_n  = \sum_{m} C_{mn} \vphi_m \,.
\label{eq:uwf}
\end{equation}
The coefficients $C_{mn}$ were obtained by solving the HFD matrix equations
\begin{equation}
  {\hat F} {\bf C}_n = \veps_n {\hat S} {\bf C}_n \,,
\label{eq:hfdeq}
\end{equation}
where ${\hat F}$ is the Dirac-Fock matrix in the joined basis
of DF and DFS orbitals of a free ion. If necessary,
an arbitrary external field
can be included in the ${\hat F}$ matrix. The matrix ${\hat S}$ is
nonorthogonal, since the DFS orbitals are not orthogonal in the usual sense.
The negative-energy DFS functions were included in the total basis set as well.
Eq.~(\ref{eq:hfdeq}) was used to generate the whole set of orthogonal
one-electron wave functions $\left\{u_n\right\}_{n = 1}^M$.

It should be noted that if even there is no external field
in Eq.~(\ref{eq:hfdeq}), the set of one-electron functions
$\left\{u_n\right\}_{n = 1}^M$ differs from the set of basis
functions $\left\{\vphi_n\right\}_{n = 1}^M$. For the occupied states,
the UDF method accounts for core-polarization effects, in contrast
to the RDF method. For the vacant states the difference is more
significant, since the DF and DFS operators are inherently different.

The many-electron wave function $\Psi_{+}(\gamma J M_J)$ with quantum numbers
$\gamma$, $J$, and $M_J$ is expanded in terms of a large set
of configuration state functions (CSFs) $\Phi_{\alpha}(J M_J)$
\begin{equation}
\Psi_{+}(\gamma J M_J) \,=\, \Lambda_{+} \Psi(\gamma J M_J) \,=\,
\sum_{\alpha} c_{\alpha} \Phi_{\alpha}(J M_J) \,.
\label{eq:Psi}
\end{equation}
The standard configuration-interaction Dirac-Fock (CIDF) method
is used to find the coefficients $c_{\alpha}$. The CSFs are
constructed from the one-electron wave functions $u_n$ (\ref{eq:uwf})
as a linear combination of Slater determinants. The set of the CSFs
is generated including all single, double, and triple excitations
into one-electron states of the positive spectrum.

\section{Negative-continuum contribution}
\label{sec:iicneg}
%
Due to some freedom 
in the choice of the wave function set
$\left\{u_n\right\}$, the positive-energy subspace
and the corresponding projector $\lambda_{+}$ Eq. (\ref{eq:lambda})
can be determined in different ways.
This freedom can be used to find the optimum many-electron wave function
$\Psiopt$ within the variational method.

The energy determined by Hamiltonian
 (\ref{eq:Hnp}) can be written as
\begin{equation}
  E = \langle \Psi \mid \Hnp \mid \Psi \rangle =
    \langle \Psi_{+} \mid \Hme \mid \Psi_{+} \rangle \,,
\qquad
    \Psi_{+} = \Lambda_{+} \Psi \,.
\end{equation}
The real orthogonal transformation (rotation) of the one-electron function
space $\left\{u_n\right\}$
modifies the wave function $\Psi_{+}$ \cite{Dalgaard}
\begin{equation}
\Psi^{\prime} = \exp(T) \Psi_{+} \,,
\end{equation}
where the operator $T$ is antihermitian ($T^{\dagger} = - T$),
\begin{equation}
  T = \sum_{n < m} E_{nm} t_{nm} \,, \qquad
  E_{nm} = a^{\dagger}_n a_m - a^{\dagger}_m a_n \,.
\end{equation}
Here $a^{\dagger}_n$ and $a_n$ are the creation and annihilation operators
of electron in the $u_n$ state. The matrix elements $t_{nm}$
can be obtained from the variational principle. Then the wave function
$\Psiopt$ satisfies the generalized Brillouin theorem \cite{Levy}
\begin{equation}
  \langle \Psiopt \mid \left[ a^{\dagger}_n a_m, \Hme \right]
    \mid \Psiopt \rangle = 0 \,.
\label{Bril}
\end{equation}
This means that the optimum wave function $\Psiopt$ is invariable
under the single excitations including negative-energy spectrum
excitations. However, this does not hold for the wave function $\Psi_{+}$.
Therefore, one should revise the calculation of the matrix element
$\langle \Psi_+ \mid A \mid  \Psi_+ \rangle$
of any one-electron operator $A$ by admixing the negative-energy spectrum
excitations to $\Psi_{+}$. This is especially important for so-called
``odd'' operators, which mix the large and small components of the
Dirac wave functions. 
The M1-transition operator $\bfT^1$ (\ref{eq:Tdip})
is just of this kind. For this reason, the negative-continuum
contribution can be significant and depends on the choice of the
one-electron basis set $\left\{u_n\right\}$ \cite{Indelicato,Derevianko}.

We consider two equivalent methods for evaluating the negative-continuum
contribution to the matrix elements of a hermitian one-electron operator $A$
with the wave functions $\Psi_{+}$. The first one is based on
the Hellman-Feynman theorem whereas the second one employs the perturbation
theory.

The space of the wave functions used to find $\Psiopt$ is invariant
under the transformation $U = \exp(iA)$, if $A$ is a one-particle operator.
Therefore, one can employ the Hellman-Feynman theorem \cite{Epstein} to
obtain the expectation value of $A$
\begin{equation}
  \overline A = \left . \frac{\partial}{\partial \mu}
    \langle \Psiopt (\mu) \mid \Hme(\mu) \mid \Psiopt (\mu) \rangle
    \right |_{\mu=0} \,,
\qquad
  \Hme (\mu) = \Hme + \mu A \,,
\end{equation}
where it is implied that $\mu A$ is included into the one-particle
Hamiltonian, $h^u(\mu)=h^u+\mu A$.
Since the wave function correction
\begin{equation}
\delta \Psi \,=\, \Psiopt - \Psi_{+} \,=\,
[1-\exp(-T)] \, \Psiopt \,\simeq\, -\sum_{n < m} E_{nm} t_{nm} \Psiopt \,
\end{equation}
accounts for single excitations only, the generalized
Brillouin theorem (\ref{Bril}) yields
\begin{equation}
  \langle \delta \Psi (\mu) \mid \Hme (\mu) \mid \Psiopt (\mu) \rangle +
  \langle \Psiopt (\mu) \mid \Hme (\mu) \mid \delta \Psi (\mu) \rangle = 0 \,
\end{equation}
and, therefore,
\begin{equation}
  \overline A = \frac{\partial}{\partial \mu} \Bigl [
    \langle \Psi_{+} (\mu) \mid \Hme(\mu) \mid \Psi_{+} (\mu) \rangle -
    \langle \delta \Psi (\mu) \mid \Hme (\mu) \mid \delta \Psi (\mu) \rangle
  \Bigr ]_{\mu=0} \,.
\end{equation}
Neglecting the second quadratic term in the equation above yields
\begin{equation}
  \overline A \simeq \frac{\partial}{\partial \mu} \Bigl [
    \langle \Psi_{+} (\mu) \mid \Hme (\mu) \mid \Psi_{+} (\mu) \rangle
  \Bigr ]_{\mu=0} \,.
\end{equation}
Thus, the negative-continuum contribution can be evaluated by means of the formula
\begin{equation}
  \Delta \overline A_{\rm neg} = \frac{\partial}{\partial \mu} \Bigl [
    \langle \Psi_{+} (\mu) \mid \Hme (\mu) \mid \Psi_{+} (\mu) \rangle
  \Bigr ]_{\mu=0} - \langle \Psi_{+} \mid A \mid \Psi_{+} \rangle\,.
\label{eq:dAneg}
\end{equation}

Alternative expression for this contribution can be obtained employing the
perturbation theory. Using the equation for the derivative of $u_n(\mu)$
\begin{equation}
 \, \left . \frac {\partial}{\partial \mu}
u_n(\mu) \right |_{\mu=0} \,=\,
\sum_{m \ne n} \, \frac{\langle u_m(0) \mid A \mid u_n(0) \rangle}
{\varepsilon_n \,-\, \varepsilon_m} \, u_m(0) \,,
\end{equation}
we obtain
\begin{equation}
\Delta \overline A_{\rm neg} \,=\, 2 \, {\sum_n}^{({\rm pos})}
\,{\sum_{m}}^{({\rm neg})} \,
\frac{\langle u_m \mid A \mid u_n \rangle}
{\varepsilon_n \,-\, \varepsilon_m} \,\,
\langle a^{+}_m \, a_n \,\Psi_{+} \mid H \mid
\Psi_{+} \rangle \,.
\label{av2}
\end{equation}
Here the indices $({\rm pos})$ and $({\rm neg})$ indicate that the
summation is carried out over the positive- and negative-energy
spectrum, respectively.

For the nondiagonal matrix elements, one can derive
\begin{equation}
\Delta  A^{i \to f}_{\rm neg} \,=\,  \frac{\partial}{\partial \mu} \, \left [
\langle \Psi^f_{+}(\mu) \mid H(\mu) \mid \Psi^i_{+}(\mu) \rangle
\right ]_{\mu=0} \,-\,
\langle \Psi^f_{+} \mid A \mid \Psi^i_{+} \rangle
\label{av3}
\end{equation}
and
\begin{eqnarray}
\Delta  A^{i \to f}_{\rm neg} &=&  {\sum_n}^{({\rm pos})}
\,{\sum_{m}}^{({\rm neg})}
\, \frac{\langle u_m \mid A \mid u_n \rangle} {\varepsilon_n
\,-\, \varepsilon_m} 
\nonumber\\
&&\times\left [ \langle a^{+}_m \, a_n
\,\Psi^f_{+} \mid H \mid \Psi^i_{+} \rangle \,+\, \langle
\Psi^f_{+} \mid H \mid a^{+}_m \, a_n \, \Psi^i_{+}
\rangle  \right] \,. \label{av4}
\end{eqnarray}
These formulas were used in our calculations of the negative-continuum
contribution to the M1-transition amplitude. It was found
that the results obtained by means of Eqs. (\ref{av3}) and (\ref{av4})
are in a perfect agreement with each other.

\section{Higher-order interelectronic-interaction corrections}
\label{sec:iicho}
%
The rigorous QED treatment of the interelectronic-interaction corrections
to the transition probabilities
can be carried out utilizing the two-time Green function
method \cite{Shabaev1}. In Ref. \cite{Indelicato04} it was done
for the $1/Z$ interelectronic-interaction corrections 
in He-like ions.
Here we perform the corresponding calculations  for B-like ions.
To simplify the derivation of formal expressions, we specify the formalism
regarding the core electrons as belonging to a redefined vacuum
(for details we refer to Refs. \cite{Shabaev1,Shabaev2}).
This leads to merging the interelectronic-interaction
corrections of order $1/Z$ with the one-loop radiative
corrections. The formulas for these corrections can easily be
obtained from the corresponding expressions for the one-loop radiative
corrections to the transition amplitude in a one-electron atom,
derived in \cite{Shabaev1}.
However, the standard electron propagator $S(\veps,\bfx,\bfy)$,
which enters the equations, must be replaced by
\begin{equation}
  \tilde{S}(\veps,\bfx,\bfy) = S(\veps,\bfx,\bfy) + 2 \pi i
    \sum_c \psi_c(\bfx) \overline{\psi}_c(\bfy) \delta(\veps-\veps_c) \,,
\label{eq:tS}
\end{equation}
where the summation runs over all occupied one-electron states
refering to the closed shells. Accordingly, the total
expression is represented by the sum of the pure QED and
interelectronic-interaction contributions,
which correspond to the first and second
terms in the right-hand side of Eq. (\ref{eq:tS}).
As a result, the $1/Z$ interelectronic-interaction correction
to the M1-transition amplitude in a B-like
ion between the initial state $a$ and the final state $b$
is
\begin{eqnarray}
  \Delta A^{\rm int}_{1M}  &=& 
- \, \sqrt{\frac{\omega}{\pi c}} \, \sqrt{3} 
\sum_c \left\{ \sum_{n \not = b}
    \frac{ \langle b c | I(0) | n c \rangle \langle n | T_M^1 | a \rangle }
    {\veps_b-\veps_n} + \sum_{n \not = a} \frac{ \langle b | T_M^1 | n \rangle
    \langle c n | I(0) | c a \rangle } {\veps_a-\veps_n} \right .
\nonumber\\
  && \left . + \sum_n \frac{ \langle b c | I(\veps_a-\veps_b) | a n \rangle
    \langle n | T_M^1 | c \rangle } {\veps_b+\veps_c-\veps_a-\veps_n}
    + \sum_n \frac{ \langle c | T_M^1 | n \rangle \langle n b | I(\veps_a-\veps_b)
    | c a \rangle} {\veps_a+\veps_c-\veps_b-\veps_n} \right .\nonumber\\
  &&
    \left . - \sum_{n \not = b} \frac{ \langle b c | I(\veps_b-\veps_c)
    | c n \rangle \langle n | T_M^1 | a \rangle } {\veps_b-\veps_n}
    - \sum_{n \not = a} \frac{ \langle b | T_M^1 | n \rangle \langle n c |
    I(\veps_a-\veps_c) | c a \rangle } {\veps_a-\veps_n} \right .
\nonumber\\
  && 
    \left . -\sum_n \frac{ \langle b c | I(\veps_a-\veps_c) | n a \rangle
    \langle n | T_M^1 | c \rangle } {\veps_b+\veps_c-\veps_a-\veps_n}
    - \sum_n \frac{ \langle c | T_M^1 | n \rangle \langle b n | I(\veps_b-\veps_c)
    | c a \rangle} {\veps_a+\veps_c-\veps_b-\veps_n} \right .
\nonumber\\
  && -
    \left .
    \frac12 \langle b | T_M^1 | a \rangle \left[
    \langle b c | I^{\prime}(\veps_b-\veps_c) | c b \rangle +
    \langle a c | I^{\prime}(\veps_a-\veps_c) | c a \rangle \right]
    \right\} \,,
\label{eq:Rexrdi}
\end{eqnarray}
where $I(\veps) =  \alpha^{\mu} \alpha^{\nu} D_{\mu\nu}(\veps)$,
 $I^{\prime}(\veps) = \rmd I(\veps) / \rmd \veps$, 
$\alpha^{\mu}=(1,\balpha)$,
and $D_{\mu\nu}(\veps)$
is the photon propagator. In the Feynman gauge it reads
\begin{equation}
  D_{\mu\nu} (\veps,\bfx-\bfy) = -4\pi g_{\mu\nu} \int\frac{\ddd k}{(2\pi)^3} \,
    \frac{\exp{(i\bfk\cdot(\bfx-\bfy))}}{\veps^2-\bfk^2+i0}\,,
\end{equation}
where $g_{\mu\nu}$ is the metric tensor.
In the Coulomb gauge we have
\begin{equation}
\begin{array}{lll}
\displaystyle
  D_{00} (\veps,\bfx-\bfy) &=&
\displaystyle
    \frac{1}{|\bfx-\bfy|} \,,
\qquad
  D_{i0}=D_{0i}=0 \,,
\qquad
  (i=1,2,3) \,,
\\
[4mm]
  D_{ij} (\veps,\bfx-\bfy) &=&
\displaystyle
 4\pi \int\frac{\ddd k}{(2\pi)^3} \,
    \frac{\exp{(i\bfk\cdot(\bfx-\bfy))}}{\veps^2-\bfk^2+i0} \,
    \left( \delta_{i,j} - \frac{k_i k_j}{\bfk^2} \right) \,,
\qquad
  (i,j=1,2,3) \,.
\end{array}
\end{equation}
In contrast to Ref. \cite{Shabaev1}, here atomic units and
the Gauss charge unit ($\alpha = e^2/\hbar c$) are used.
Expression (\ref{eq:Rexrdi}) incorporates the 
Coulomb-Breit part, which was taken into account
by the CI method, together with terms of higher order in $\az$,
the so-called frequency-dependent correction. Specifying the
operator $I(\veps)$ within the Coulomb gauge and setting
$\veps=0$ in Eq. (\ref{eq:Rexrdi}) yields the Coulomb-Breit interaction.
In this way we can exclude the part, which has already been taken
into account by the CI method, and obtain the frequency-dependent
correction of order $1/Z$ as
\begin{eqnarray}
 \Delta A^{\rm freq}_{1M}  &=&
 \, \sqrt{\frac{\omega}{\pi c}} \, \sqrt{3}
\sum_c \left \{ \sum_{n \not = b} \frac{\langle b c |
    \Delta \IC(\veps_b-\veps_c) | c n \rangle \langle n | T_M^1 | a \rangle }
    {\veps_b-\veps_n}\right .
\nonumber\\
&& + \sum_{n \not = a} \frac { \langle b | T_M^1 | n \rangle
    \langle n c | \Delta \IC(\veps_a-\veps_c) | c a \rangle} {\veps_a-\veps_n}
\nonumber\\
  && +
    \sum_n \frac{ \langle b c | \Delta \IC(\veps_a-\veps_c) | n a \rangle
    \langle n | T_M^1 | c \rangle} {\veps_b+\veps_c-\veps_a-\veps_n} + \sum_n
    \frac{ \langle c | T_M^1 | n \rangle \langle b n | \Delta \IC(\veps_b-\veps_c)
    | c a \rangle} {\veps_a+\veps_c-\veps_b-\veps_n}
\nonumber\\
  && -
    \sum_n \frac{ \langle b c | \Delta \IC(\veps_a-\veps_b) | a n \rangle
    \langle n | T_M^1 | c \rangle } {\veps_b+\veps_c-\veps_a-\veps_n} - \sum_n
    \frac{ \langle c | T_M^1 | n \rangle \langle n b | \Delta \IC(\veps_a-\veps_b)
    | c a\rangle} {\veps_a+\veps_c-\veps_b-\veps_n}
\nonumber\\
  && + \left .
    \frac12 \langle b | T_M^1 | a \rangle \left[ \langle b c |
    \IC^{\prime}(\veps_b-\veps_c) | c b \rangle + \langle a c |
    \IC^{\prime}(\veps_a-\veps_c) | c a \rangle \right] \, \right\} \,,
\end{eqnarray}
where $\Delta \IC(\veps_a-\veps_b) = \IC(\veps_a-\veps_b) - \IC(0)$
and the subscript ``C'' refers to the Coulomb gauge.

It should be noted that the total $1/Z$
interelectronic-interaction correction given by equation
(\ref{eq:Rexrdi})  is gauge independent. This has been
confirmed in our calculations to a very high accuracy.
The calculations were performed employing the B-spline method for
the Dirac equation \cite{joh88}.

\section{QED correction}
\label{sec:qed}
%
QED effects modify the transition probability via
the matrix element of the transition operator and via the transition energy.
Since we employ the experimental value for the transition energy,
we have to consider the QED effect on the transition amplitude only.

The lowest-order QED correction to the M1-transition amplitude
can be derived by correcting the operator of the atomic
magnetic moment for the anomalous magnetic moment of a free electron.
In the nonrelativistic limit it yields
\begin{equation}
\bmu_{\rm nr} \rightarrow \bmu_a = - \mub \left [ \bfL + 2 (1 + \kappa_e) \bfS \right ]
       = \bmu_{\rm nr} + \delta \bmu_a \,,
\end{equation}
where
\begin{eqnarray}
  \delta \bmu_a = - 2 \mub \kappa_e \bfS \,,
\end{eqnarray}
\begin{eqnarray}
  \kappa_e = \left [ \frac{\alpha}{2\pi} - 0.328\,478\,965\ldots \,
    \left ( \frac{\alpha}{\pi} \right )^2 + \cdots \right ] \,.
\end{eqnarray}
With the aid of the identity
\begin{equation}
  \brafr \bfJ \ketir = \brafr (\bfL + \bfS) \ketir =
  \delta_{\jjf,\jji} \sqrt{\jji(\jji+1)(2\jji+1)} \,,
\end{equation}
one can easily find for the fine-structure level transition $(\Delta J = \pm 1)$
\begin{equation}
  \brafr \delta \bmu_a \ketir =
  2 \kappa_e \brafr \bmu_{\rm nr} \ketir \,.
\end{equation}
Therefore, the QED correction to the M1-transition probability is given by
\begin{equation}
  \dwifqed = \frac{4\omega^3}{3c^3}\frac{1}{2\jji+1} \left(
    \left| \brafr (\bmu_{\rm nr} + \delta \bmu_a) \ketir \right|^2 -
    \left| \brafr \bmu_{\rm nr} \ketir \right|^2 \right) \,,
\end{equation}
which yields
\begin{eqnarray}\label{qqed_finn}
  \dwifqed \simeq
     4 \kappa_e \frac{4\omega^3}{3c^3}\frac{1}{2\jji+1}
    \left| \brafr \bmu_{\rm nr} \ketir \right|^2 \simeq 4 \kappa_e \wifnr \,.
\end{eqnarray} 
QED corrections, which are not accounted for by this formula,
are suppressed by a small factor $(\alpha Z)^2$.

\section{Results and discussion}
\label{sec:results}

The individual contributions to the M1-transition probabilities and 
the corresponding lifetimes for B-like and Be-like ions are
presented in Tables~\ref{tab:WB}~and~\ref{tab:WBe}, respectively. 
Due to the smallness of the E2 transition, which is also allowed,
the lifetimes are essentially determined by the M1 transition.
In case of B-like ions, the experimental values of the transition energy 
were taken from  Ref. \cite{Edlen} for S$^{11+}$, Cl$^{12+}$, K$^{14+}$,
Ti$^{17+}$ and from Ref. \cite{Draganic} for Ar$^{13+}$.
As one can see from Table~\ref{tab:WB} the interelectronic-interaction
correction $\dwci$ turns out to be relatively small due to the
smallness of the factor $(\az)^2/Z$. The most important contributions
are given by the relativistic correction $\dwd$ and by the QED
correction $\dwqed$.
For Be-like ions, the transition energies were taken
from Ref. \cite{EdlenBe} for S$^{12+}$, Cl$^{13+}$, K$^{15+}$, Ti$^{18+}$
and from Ref. \cite{Draganic} for Ar$^{14+}$. In this case the
interelectronic-interaction correction $\dwci$ provides an essential
contribution to the total value of the transition probability. This 
is due to a strong mixing of the two terms $^3P_1$ and $^1P_1$.
Except for Ar$^{13+}$ and   Ar$^{14+}$, the uncertainties of the
total transition probabilities are mainly determined by the
experimental uncertainties of the transition energy. For argon ions,
the uncertainty comes mainly from uncalculated higher-order QED corrections.

In  Table~\ref{tab:WBcomp}, our results for the lifetime of the $(1s^2
2s^2 2p)\,^2P_{3/2}$ state are compared with other calculations and with
experiment. It should be noted that the QED correction was taken into account
in Refs.~\cite{fisch83,Johnson} and in the present work only.
Besides, different values of the transition energy $\omega$, indicated in
Table~\ref{tab:WBcomp}, were used in the different calculations. Since
the M1-transition probability $W$ scales as $\omega^3$, a small
deviation in $\omega$ can change $W$ significantly. For this reason,
we recalculated the results of Cheng {\it et al.} \cite{cheng79} and
Froese Fischer \cite{fisch83} for the $(1s^2 2s^2 2p)\,^2P_{3/2}$
state in B-like ions for those transition energies we have employed in
our calculations. Table~\ref{tab:WBcompe} presents these 
values with ($\tau$ \cite{fisch83}) and without ($\tau^0$
\cite{cheng79}) the anomalous
magnetic moment correction and the corresponding values ($\tpres$ and
$\tpres^0$) obtained in this work. As one can see from
the table, there is an excellent agreement between
our ``non-QED'' results ($\tpres^0$) and those from Ref. \cite{cheng79}
($\tau^0$). There is also a good agreement between our total
results ($\tpres$) and those from Ref. \cite{fisch83} ($\tau$). The comparison
of our theoretical results with the experimental data shows generally a
good agreement as well. However, in case of Ar$^{13+}$ there is a
discrepancy between our $^2P_{3/2}$ lifetime value $9.538(2)$ ms and
the most accurate experimental value $9.573(4)(5)$ ms
\cite{Lapierre,Lapierre1}.

Table~\ref{tab:WBecomp} shows a fair agreement of our results
for the lifetime of the $(1s^2 2s 2p)\,^3P_{2}$ state in Be-like ions
with corresponding results obtained by other authors and with experimental
data. We note that the QED correction has not been considered in the previous
calculations cited in the table.

In conclusion, we have evaluated the magnetic-dipole transition probabilities
between the fine-structure levels
$(1s^2 2s^2 2p) \, ^2P_{1/2} \,-\, ^2P_{3/2}$ for B-like
ions and $(1s^2 2s 2p) \, ^3P_{1} \,-\, ^3P_{2}$ for
Be-like ions. The relativistic, interelectronic-interaction, and
radiative corrections to the transition probability  have been considered.
Except for a recent high-precision lifetime measurement on Ar$^{13+}$
\cite{Lapierre,Lapierre1} with an accuracy level on the order of 0.1\%, 
most experimental results have large error
bars greater than 1.5\% and, within these error bars, most of them are in a 
fair agreement with our theoretical
predictions. In case of Ar$^{13+}$, the disagreement of our prediction 
with the high-precision experimental value
amounts to 0.37\% of the total transition probability, less than 
the value of the corresponding QED correction.
At present we have no explanation for this discrepancy.

\acknowledgments
%

Valuable conversations with O. Yu. Andreev are gratefully acknowledged.
This work was supported in part by RFBR (Grant No.
04-02-17574), INTAS-GSI (Grant No. 03-54-3604), the Russian Ministry
of Education. D.A.G. acknowledges financial support from the
foundation ``Dynasty''. A.V.V. and G.P. acknowledge financial support
from the GSI F+E program, DFG, and BMBF. The work of A.V.V. was also
supported by DAAD and ``Dynasty'' foundation.
%

%
\newpage
\begin{table}
\begin{center}
\caption{The decay rates $W$ [s$^{-1}$] of the magnetic-dipole
transition $(1s^2 2s^2 2p)\ ^2P_{1/2} -\ ^2P_{3/2}$ and the
lifetimes $\tau$ [ms] of the $(1s^2 2s^2 2p)\ ^2P_{3/2}$ state
in B-like ions. Numbers in the parentheses give the estimated error.
\label{tab:WB}}
\tabcolsep10pt
\begin{tabular}{lrrrrr}\\[-4mm]
\hline \\[-4mm]
                      &S$^{11+}$\hsp{3pt} &Cl$^{12+}$\hsp{3pt} & Ar$^{13+}$ &K$^{14+}$ &Ti$^{17+}$\hsp{2pt} \\[1mm]
\hline \\[-4mm]
Energy [cm$^{-1}$]    & 13135(1)\hsp{3pt} & 17408(20)\hsp{3pt} &22656.22(1) &29006(25) & 56243(4)\hsp{2pt}  \\[1mm]
\hline \\[-4mm]
$W_{\rm nr}$          & 20.37538\hsp{3pt} & 47.43068\hsp{3pt}  & 104.56308  & 219.4222 & 1599.635\hsp{2pt}  \\
$\Delta W_{\rm D}$    & -0.03542\hsp{3pt} & -0.09302\hsp{3pt}  &  -0.23145  &  -0.5436 &   -5.355\hsp{2pt}  \\
$\Delta W_{\rm CI}$   &  0.00637\hsp{3pt} &  0.01586\hsp{3pt}  &   0.03723  &   0.0802 &    0.597\hsp{2pt}  \\
$\Delta W_{\rm neg}$  & -0.00159\hsp{3pt} & -0.00396\hsp{3pt}  &  -0.00929  &  -0.0206 &   -0.176\hsp{2pt}  \\
$\Delta W_{\rm QED}$  &  0.09451\hsp{3pt} &  0.22001\hsp{3pt}  &   0.48502  &   1.0178 &    7.420\hsp{2pt}  \\
$\Delta W_{\rm freq}$ &  0.00007\hsp{3pt} &  0.00019\hsp{3pt}  &   0.00049  &   0.0012 &    0.013\hsp{2pt}  \\[1mm]
\hline \\[-4mm]
$W_{\rm total}$       & 20.439(5)         & 47.57(16)          & 104.85(2)\hsp{3pt}
                                                                            & 220.0(6)\hsp{4pt}
                                                                                       & 1602.1(5)          \\
$\tau_{\rm total}$    & 48.93(1)          & 21.02(7)           & 9.538(2)   & 4.546(12)& 0.6242(2)          \\[-0mm]
\hline
\end{tabular}
\end{center}
\end{table}
\begin{table}
\begin{center}
\caption{The decay rates $W$ [s$^{-1}$] of the magnetic-dipole
transition $(1s^2 2s 2p)\ ^3P_1 -\ ^3P_2$ and the
lifetimes $\tau$ [ms] of the $(1s^2 2s 2p)\ ^3P_2$ state
in Be-like ions. Numbers in the parentheses give the estimated error.
\label{tab:WBe}}
\tabcolsep10pt
\begin{tabular}{lrrrrr}\\[-4mm]
\hline \\[-4mm]
                      & S$^{12+}$ & Cl$^{13+}$\hsp{3pt} & Ar$^{14+}$ & K$^{15+}$ & Ti$^{18+}$        \\[1mm]
\hline \\[-4mm]
Energy [cm$^{-1}$]    & 9712(14)  & 12913(16)\hsp{3pt}  & 16819.36(1)& 21571(20) & 42638(4)          \\[1mm]
\hline \\[-4mm]
$W_{\rm nr}$          & 12.35488  & 29.03947\hsp{3pt}   & 64.17056   &135.36899  &1045.4311          \\
$\Delta W_{\rm D}$    & -0.02017  & -0.05389\hsp{3pt}   & -0.13242   & -0.31247  &  -3.2611          \\
$\Delta W_{\rm CI}$   & -0.01302  & -0.04909\hsp{3pt}   & -0.16457   & -0.50484  & -10.0481          \\
$\Delta W_{\rm neg}$  & -0.00053  & -0.00133\hsp{3pt}   & -0.00313   & -0.00704  &  -0.0649          \\
$\Delta W_{\rm QED}$  &  0.05731  &  0.13470\hsp{3pt}   &  0.29766   &  0.62792  &   4.8493          \\[1mm]
\hline \\[-4mm]
$W_{\rm total}$       & 12.38(5)\hsp{3pt}
                                  & 29.07(11)           & 64.17(1)\hsp{3pt}
                                                                     &135.2(4)\hsp{8pt}
                                                                                 &1036.9(4)\hsp{2pt} \\
$\tau_{\rm total}$    & 80.79(33) & 34.40(13)           & 15.584(2)  & 7.398(22) & 0.9645(4)         \\[-0mm]
\hline
\end{tabular}
\end{center}
\end{table}
\begin{table}
\begin{center}
\caption{ The lifetimes of the $(1s^2 2s^2 2p)\ ^2P_{3/2}$ level
in B-like ions calculated in this work with ($\tau_{\rm pres}$)
and without ($\tau^0_{\rm pres}$) the QED correction are compared
with previous calculations ($\tau_{\rm theor}$) and experiment
($\tau_{\rm exp}$). The lifetime values are given in [ms].
The values of the transition energy [Energy]
are presented in [cm$^{-1}$]. Numbers in the parentheses give the estimated error.
\label{tab:WBcomp}}
{\footnotesize
\tabcolsep8pt
\begin{tabular}{lrrlrl}\\[-8mm]
\hline \\[-8mm]
Ions & $\tau^0_{\rm pres}$ & $\tau_{\rm pres}$[Energy] & $\tau_{\rm theor}$[Energy]
                                                       &~~Method \& Ref.&            $\tau_{\rm exp}$ \& Ref.           \\[1mm]
\hline \\[-8mm]
 S$^{11+}$ & 49.16& 48.93(1) [13135]  & 47.35 [13300]  & MCDF \cite{cheng79}   \hsp{6pt} &                              \\ [-3mm]
           &      &                   & 49.07 [13115]  & MCBP \cite{fisch83}   \hsp{6pt} &                              \\ [-3mm]
           &      &                   & 49.33 [13144]  & MCDF \cite{ver87}     \hsp{6pt} &                              \\ [-3mm]
           &      &                   & 49.07 [13136]  & SS   \cite{gala98}    \hsp{6pt} &                              \\ [-3mm]
           &      &                   & 49.26 [13122]  & MRCI \cite{Koc}       \hsp{6pt} &                              \\ [-3mm]
           &      &                   & 49.60          & RQDO \cite{cha01}     \hsp{6pt} &                              \\[1mm]
Cl$^{12+}$ & 21.12& 21.02(7) [17408]  & 20.55 [17565]  & MCDF \cite{cheng79}   \hsp{6pt} & 21.2(6)\cite{Trabert2}       \\ [-3mm]
           &      &                   & 21.02 [17400]  & MCBP \cite{fisch83}   \hsp{6pt} & 21.1(5)\cite{Trabert2}       \\ [-3mm]
           &      &                   & 21.19 [17421]  & MCDF \cite{ver87}     \hsp{6pt} &                              \\ [-3mm]
           &      &                   & 21.08 [17410]  & SS   \cite{gala98}    \hsp{6pt} &                              \\ [-3mm]
           &      &                   & 21.19 [17386]  & MRCI \cite{Koc}       \hsp{6pt} &                              \\ [-3mm]
           &      &                   & 21.13          & RQDO \cite{cha01}     \hsp{6pt} &                              \\[1mm]
Ar$^{13+}$ & 9.582& 9.538(2) [22656]  & 9.407 [22795]  & MCDF \cite{cheng79}   \hsp{6pt} & 8.7(5)\cite{ar13}            \\ [-3mm]
           &      &                   & 9.515 [22660]  & MCBP \cite{fisch83}   \hsp{6pt} & 9.12(18)\cite{Moehs}         \\ [-3mm]
           &      &                   & 9.618 [22666]  & MCDF \cite{ver87}     \hsp{6pt} & 9.70(15)\cite{Trabert1}      \\ [-3mm]
           &      &                   & 9.569 [22653]  & SS   \cite{gala98}    \hsp{6pt} & 9.573(4)(5)\cite{Lapierre1} \\ [-3mm]
           &      &                   & 9.588 [22657]  & RQDO \cite{cha01}     \hsp{6pt} &                              \\[-3mm]
           &      &                   & 9.606 [22636]  & MCDF \cite{Fritzsche} \hsp{6pt} &                              \\[-3mm]
           &      &                   & 9.615 [22619]  & MRCI \cite{Koc}       \hsp{6pt} &                              \\[-3mm]
           &      &                   & 9.534 [22658]  &      \cite{Johnson}   \hsp{6pt} &                              \\[1mm]
K$^{14+}$  & 4.567& 4.546(12) [29006] & 4.509 [29129]  & MCDF \cite{cheng79}   \hsp{6pt} & 4.47(10)\cite{k15}           \\ [-3mm]
           &      &                   & 4.521 [29044]  & MCBP \cite{fisch83}   \hsp{6pt} &                              \\ [-3mm]
           &      &                   & 4.583 [29019]  & MCDF \cite{ver87}     \hsp{6pt} &                              \\ [-3mm]
           &      &                   & 4.558 [29004]  & SS   \cite{gala98}    \hsp{6pt} &                              \\ [-3mm]
           &      &                   & 4.587 [28960]  & MRCI \cite{Koc}       \hsp{6pt} &                              \\ [-3mm]
           &      &                   & 4.577          & RQDO \cite{cha01}     \hsp{6pt} &                              \\[1mm]
Ti$^{17+}$ &0.6271& 0.6242(2) [56243] & 0.6254 [56275] & MCDF \cite{cheng79}   \hsp{6pt} & 0.627(10)\cite{ti17}         \\ [-3mm]
           &      &                   & 0.6150 [56465] & MCBP \cite{fisch83}   \hsp{6pt} &                              \\ [-3mm]
           &      &                   & 0.6290 [56258] & MCDF \cite{ver87}     \hsp{6pt} &                              \\ [-3mm]
           &      &                   & 0.6254 [56240] & SS   \cite{gala98}    \hsp{6pt} &                              \\ [-3mm]
           &      &                   & 0.6289 [56166] & MRCI \cite{Koc}       \hsp{6pt} &                              \\ [-3mm]
           &      &                   & 0.6270         & RQDO \cite{cha01}     \hsp{6pt} &                              \\ [1mm]
\hline \\[-8mm]
\multicolumn{6}{l}{MCDF - multiconfiguration Dirac-Fock method}\\ [-3mm]
\multicolumn{6}{l}{MCBP - multiconfiguration Breit-Pauli method}\\ [-3mm]
\multicolumn{6}{l}{SS~~~~~~ - SUPERSTRUCTURE program}\\ [-3mm]
\multicolumn{6}{l}{MRCI - multireference relativistic configuration interaction method}\\ [-3mm]
\multicolumn{6}{l}{RQDO - relativistic quantum defect orbital method}\\ [0mm]
\end{tabular}
}
\end{center}
\end{table}
\begin{table}
\begin{center}
\small \caption{The lifetimes of the $(1s^2 2s^2 2p)\ ^2P_{3/2}$ level
in B-like ions calculated in this work with ($\tau_{\rm pres}$)
and without ($\tau^0_{\rm pres}$) the QED correction are compared
with previous theoretical results, recalculated to the transition energy
(Energy[cm$^{-1}$]) employed in this paper. The lifetime values are given in [ms].
 \label{tab:WBcompe}}
{\small
\tabcolsep12pt
\begin{tabular}{lllclc}
\\[-4mm] \hline \\[-8mm]
  Ions     & Energy & $\tau^0_{\rm pres}$ & $\tau^0$ (Ref. \cite{cheng79})
      & $\tau_{\rm pres}$ & $\tau$ (Ref. \cite{fisch83}) \\[1mm]
\hline \\[-8mm]
S$^{11+}$  & 13135  & 49.16               & 49.16                   & 48.93             &  48.85     \\[-3mm]
Cl$^{12+}$ & 17408  & 21.12               & 21.11                   & 21.02             &  20.99     \\[-3mm]
Ar$^{13+}$ & 22656  & 9.582               & 9.581                   & 9.538             &  9.520     \\[-3mm]
 K$^{14+}$ & 29006  & 4.567               & 4.567                   & 4.546             &  4.539     \\[-3mm]
Ti$^{17+}$ & 56243  & 0.6271              & 0.6265                  & 0.6242            &  0.6223    \\[ 0mm]
\hline \\[-20mm]
\end{tabular}
}
\end{center}
\end{table}
\begin{table}
\begin{center}
\caption{The lifetimes of the $(1s^2 2s 2p)\ ^3P_2$ level
in Be-like ions calculated in this work with ($\tau_{\rm pres}$)
and without ($\tau^0_{\rm pres}$) the QED correction are compared
with previous calculations ($\tau_{\rm theor}$) and experiment
($\tau_{\rm exp}$). The lifetime values are given in [ms].
The values of the transition energy [Energy]
are presented in [cm$^{-1}$]. Numbers in the parentheses give the estimated error.
\label{tab:WBecomp}}
{\small
\tabcolsep8pt
\begin{tabular}{lrrlrl}
\\[-2mm]
\hline \\[-8mm]
Ions & $\tau^0_{\rm pres}$ & $\tau_{\rm pres} $[Energy] & $\tau_{\rm theor}$ [Energy]
                                                           &~~Method \& Ref.       & $\tau_{\rm exp}$ \& Ref.\\[1mm]
\hline \\[-8mm]
 S$^{12+}$ & 81.16& 80.79(33)     [9712]  & 83.3 \ [9743]  & SHF  \cite{Kaufman}   &                  \\[-3mm]
           &      &                       & 80.65  [9720]  & MBPT \cite{Safronova} &                  \\[ 1mm]
Cl$^{13+}$ & 34.56& 34.40(13)    [12913]  & 35.7 \ [12893] & SHF  \cite{Kaufman}   &                   \\[-3mm]
           &      &                       & 34.60  [12903] & MBPT \cite{Safronova} &                   \\[ 1mm]
Ar$^{14+}$ & 15.66& 15.584(2)    [16819]  & 16.31  [16818] & MCHF \cite{Glass}     & 15.0(7)\cite{Back}  \\[-3mm]
           &      &                       & 16.1 \ [16824] & SHF  \cite{Kaufman}   & 13.4(7)\cite{Moehs} \\[-3mm]
           &      &                       & 15.63  [16834] & MBPT \cite{Safronova} & 15.0(8)\cite{Trabert1} \\[-3mm]
           &      &                       & 15.76  [16782] & MCDF \cite{Fritzsche} &                     \\[1mm]
K$^{15+}$  & 7.432& 7.398(22)    [21571]  & 7.63 \ [21575] & SHF  \cite{Kaufman}   & 7.6(5)\cite{k15}   \\[-3mm]
           &      &                       & 7.353  [21633] & MBPT \cite{Safronova} &                    \\[1mm]
Ti$^{18+}$ &0.9689& 0.9645(4)    [42638]  & 0.990 \ [42653]& SHF \cite{Kaufman}    &                     \\[-3mm]
           &      &                       & 0.9615  [42651]& MBPT \cite{Safronova} &                     \\
\hline \\[-8mm]
\multicolumn{6}{l}{SHF - scaled Hartree-Fock method}\\ [-2mm]
\multicolumn{6}{l}{MBPT - many-body perturbation theory}\\ [-2mm]
\multicolumn{6}{l}{MCHF - multiconfiguration Hartree-Fock method}\\ [-2mm]
\multicolumn{6}{l}{MCDF - multiconfiguration Dirac-Fock method}
\end{tabular}
}
\end{center}
\end{table}
%

\begin{thebibliography}{99}
\clearpage
\bibitem{Back}
T. V. Back, H. S. Margolis, P. K. Oxley, J. D. Silver,
and E. G. Myers, Hyperfine Int. {\bf 114}, 203 (1998).
%
\bibitem{Moehs}
D. P. Moehs and D. A. Church, Phys. Rev. A {\bf 58}, 1111 (1998).
%
\bibitem{ti17}
E. Tr\"abert, G. Gwinner, A. Wolf, X. Tordoir,
and A. G. Calamai, Phys. Lett. A {\bf 264}, 311 (1999).
%
\bibitem{Trabert1}
E. Tr\"abert, P. Beiersdorfer, S. B. Utter,
G. V. Brown, H. Chen, C. L. Harris, P. A. Neill,
D. W. Savin, and A. J. Smith, Astrophys. J. {\bf 541}, 506 (2000).
%
\bibitem{k15}
E. Tr\"abert, P. Beiersdorfer, G. V. Brown,
H. Chen, E. H. Pinnington, and D. B. Thorn,
Phys. Rev. A {\bf 64}, 034501 (2001).
%
\bibitem{Trabert2}
E. Tr\"abert, P. Beiersdorfer, G. Gwinner,
E. H. Pinnington, and A. Wolf, Phys. Rev. A {\bf 66}, 052507 (2002).
%
%
\bibitem{Lapierre}
J. R. Crespo L\'opez-Urrutia, A. N. Artemyev, J. Braun, G. Brenner,
H. Bruhns, I. N. Dragani\'c, A. J. Gonz\'alez Mart\'inez,
A. Lapierre, V. Mironov, J. Scofield, R. Soria Orts, H. Tawara,
M. Trinczek, I. I. Tupitsyn, and J. Ullrich,
Nucl. Instr. Meth. Phys. Res. B {\bf 235}, 85 (2005).
%
\bibitem{Lapierre1}
A. Lapierre, U. D. Jentschura,
J. R. Crespo L\'opez-Urrutia, J. Braun, G. Brenner,
H. Bruhns, D. Fischer, A. J. Gonz\'alez Mart\'inez, Z. Harman,
W. R. Johnson, C. H. Keitel, V. Mironov, C. J. Osborne, G. Sikler,
R. Soria Orts, H. Tawara, I. I. Tupitsyn, J. Ullrich, and A. Volotka,
accepted in Phys. Rev. Lett.
%
\bibitem{cheng79}
K. T. Cheng, Y.-K. Kim, and J. P. Desclaux,
At. Data Nucl. Data Tables {\bf 24}, 111 (1979).
%
\bibitem{fisch83}
C. F. Fischer, J. Phys. B {\bf 16}, 157 (1983).
%
\bibitem{cha01}
E. Charro, S. L\'opez-Ferrero, and I. Mart\'in,
J. Phys. B {\bf 34}, 4243 (2001).
%
\bibitem{Sobelman}
I. I. Sobelman, {\it Atomic Spectra and Radiative Transitions},
Springer, New York, 1979.
%
\bibitem{Sucher}
J. Sucher, Phys. Rev. A {\bf 22}, 348 (1980).
%
\bibitem{Mittleman}
M. H. Mittleman, Phys. Rev. A {\bf 24}, 1167 (1981).
%
\bibitem{gla04}
D. A. Glazov, V. M. Shabaev, I. I. Tupitsyn, A. V. Volotka, V. A. Yerokhin,
G. Plunien, and G. Soff, Phys. Rev. A {\bf 70}, 062104 (2004).
%
\bibitem{Indelicato}
P. Indelicato, Phys. Rev. Lett. {\bf 77}, 3323 (1996).
%
\bibitem{Derevianko}
A. Derevianko, I. M. Savukov, W. R. Johnson, and
D. R. Plante, Phys. Rev. A {\bf 58}, 4453 (1998).
%
\bibitem{Shabaev1}
V. M. Shabaev, Phys. Rep. {\bf 356}, 119 (2002).
%
\bibitem{Grant1}
I. P. Grant, J. Phys. B {\bf 7}, 1458 (1974).
%
\bibitem{Varshalovich}
D. A. Varshalovich, A. N. Moskalev, V. K. Khersonskii,
{\it Quantum Theory of Angular Momentum}, World Scientific,
Singapore, 1988.
%
\bibitem{Grant2}
I. P. Grant, Advances in Physics {\bf 19}, 747 (1970).
%
\bibitem{Bratsev}
V. F. Bratzev, G. B. Deyneka, and I. I. Tupitsyn,
Izv. Akad. Nauk SSSR {\bf 41}, 2655 (1977)
[Bull. Acad. Sci. USSR, Phys. Ser. {\bf 41}, 173 (1977)].
%
\bibitem{Gruzdev}
P. F. Gruzdev, G. S. Soloveva, and A. I. Sherstyuk,
Opt. Spektrosk. {\bf 42}, 1198 (1977)
[Opt. Spectrosc. {\bf 42}, 690 (1977)].
%
\bibitem{Dalgaard} E. Dalgaard and P. J\o rgensen, J. Chem. Phys.
                   {\bf 69}, 3833 (1978).
%
\bibitem{Levy} B. Levy and G. Berthier, Int. J. Quantum. Chem.
                   {\bf 2}, 307 (1968).
%
\bibitem{Epstein}
S. T. Epstein, {\it Variation Method in Quantum Chemistry},
Academic Press, New York, 1974.
%
\bibitem{Indelicato04}
P. Indelicato, V. M. Shabaev, and A. V. Volotka,
Phys. Rev. A {\bf 69}, 062506 (2004).
%
\bibitem{Shabaev2}
M. B. Shabaeva and V. M. Shabaev,
Phys. Rev. A {\bf 52}, 2811 (1995).
%
\bibitem{joh88}
W. R. Johnson, S. A. Blundell, and J. Sapirstein,
Phys. Rev. A {\bf 37}, 307 (1988).
%
\bibitem{Edlen}
B. Edl\'en, Phys. Scripta {\bf 28}, 483 (1983).
%
\bibitem{Draganic}
I. Dragani\'c, J. R. Crespo L\'opez-Urrutia, R. DuBois,
S. Fritzsche, V. M. Shabaev, R. Soria Orts, I. I. Tupitsyn,
Y. Zou, and J. Ullrich, Phys. Rev. Lett. {\bf 91}, 183001 (2003).
%
\bibitem{EdlenBe}
B. Edl\'en, Phys. Scripta {\bf 28}, 51 (1983).
%
\bibitem{ver87}
T. R. Verhey, B. P. Das, and W. F. Perger,
J. Phys. B {\bf 20}, 3639 (1987).
%
\bibitem{gala98}
M. E. Galav\'is, C. Mendoza, and C. J. Zeippen,
Astron. Astrophys. Suppl. Ser. {\bf 131}, 499 (1998).
%
\bibitem{Koc}
K. Koc, J. Phys. B {\bf 36}, L93 (2003).
%
\bibitem{Fritzsche}
C. Z. Dong, S. Fritzsche, B. Fricke, and W.-D. Sepp,
Phys. Scripta {\bf T92}, 294 (2001).
%
\bibitem{Johnson}
W. R. Johnson, private communication.
%
\bibitem{ar13}
F. G. Serpa, J. D. Gillaspy, and E. Tr\"abert,
J. Phys. B {\bf 31}, 3345 (1998).
%
\bibitem{Kaufman}
V. Kaufman and J. Sugar, J. Phys. Chem. Ref. Data {\bf 15},
321 (1986).
%
\bibitem{Safronova}
U. I. Safronova, W. R. Johnson, and A. Derevianko,
Phys. Scr. {\bf 60}, 46 (1999).
%
\bibitem{Glass}
R. Glass, Astrophys. Space Sci. {\bf 91}, 417 (1983).
\end{thebibliography}
\end{document}